\documentclass[%
superscriptaddress,
preprint,
showpacs,preprintnumbers,
amsmath,amssymb,
aps,
]{revtex4-1}

\usepackage{amsthm}
\usepackage{color}
\usepackage{graphicx}
\usepackage{dcolumn}
\usepackage{bm}

\newtheorem{remark}{Remark}
\newtheorem{proposition}{Proposition}
\newtheorem{lemma}{Lemma}

\begin {document}

\title{Distributional behavior of time averages of non-$L^1$ observables
 in one-dimensional intermittent maps with infinite invariant measures}

\author{Takuma Akimoto}
\email{akimoto@z8.keio.jp}
\affiliation{%
  Department of Mechanical Engineering, Keio University, Yokohama, 223-8522, Japan
}%

\author{Soya Shinkai}
\affiliation{%
  Research Center for the Mathematics on Chromatin Live Dynamics (RcMcD), Hiroshima University, 739-8530, Japan
}%

\author{Yoji Aizawa}
\affiliation{%
  Department of Applied Physics, Advanced School of Science and Engineering, Waseda University, Tokyo 169-8555, Japan
}%


\date{\today}

\begin{abstract}
In infinite ergodic theory, two
    distributional limit theorems are well-known. One is characterized by the Mittag-Leffler
    distribution for time averages of $L^1(m)$ functions, i.e., integrable functions with respect to an 
    infinite invariant measure. The other is characterized by 
    the generalized arc-sine distribution for time averages of
    non-$L^1(m)$ functions. Here, we provide another distributional behavior of time averages 
    of non-$L^1(m)$ functions in one-dimensional intermittent maps where each has an indifferent fixed point 
    and an infinite invariant measure. Observation functions considered here are
    non-$L^1(m)$ functions which vanish at the indifferent fixed
    point. 
    We call this class of observation functions  {\it weak non-$L^1(m)$ function}. 
    Our main result represents a first step toward a third distributional limit theorem, i.e., a distributional limit theorem for this class of
    observables, in infinite ergodic theory. To prove our proposition, we propose 
    a stochastic process induced by a renewal process to mimic a Birkoff sum of a weak non-$L^1(m)$ 
    function in the one-dimensional intermittent maps.
\end{abstract}

\maketitle

\maketitle

\section {Introduction}

In statistical physics, {many observables 
result from  time averages of the microscopic observation functions.} 
Ergodic theory plays an important role in providing the asymptotic behavior of time-averaged observables in 
dynamical systems. Trajectories in chaotic dynamical systems cannot be predicted due to the sensitivity dependence 
of initial conditions. However, with the aid of the unpredictability, trajectories can be regarded as a stochastic process. Then,  
one can introduce a measure in dynamical systems. In fact, an invariant measure 
 characterizes chaotic orbits. 
Birkhoff's ergodic theorem tells us that {time averages of an observation function
converge to a constant for almost all initial conditions if the observation function is integrable 
with respect to an absolutely continuous invariant measure \cite{Birkhoff1931}. }
On the other hand, when an invariant measure cannot be normalized (infinite measure), the asymptotic behavior of time-averaged 
observables is completely different from that stated by the Birkhoff's ergodic theorem.  In infinite measure systems (infinite ergodic 
theory), {one of 
the most striking points is that a time-averaged observable does not converge to a constant but converges} 
in distribution \cite{Aaronson1981,Aaronson1997,Thaler1998,Thaler2002,Akimoto2008}.

 In infinite ergodic theory, two different distributional limit theorems for time averages have been known. 
Distribution of time averages of an $L^1(m)$ function, which is an integrable function with respect to an invariant 
measure $m$, follows the Mittag-Leffler distribution \cite{Aaronson1981,Aaronson1997}. 
This distributional limit theorem is based on Darling-Kac theorem 
 in stochastic processes \cite{Darling1957}. The other distributional limit theorem states that time averages of a non-$L^1(m)$ function
converges in distribution to the generalized arc-sin distribution \cite{Aizawa1989,Thaler1998,Thaler2002,TZ2006,Akimoto2008}, which 
is based on Dynkin-Lamperti's generalized arc-sine law \cite{Lamperti1958,Dynkin1961}. In infinite ergodic theory, 
it is important to determine the distribution of time averages for arbitrary observation functions as well as arbitrary { ensembles of 
initial points.}  
Recently, one of the authors has shown 
that the distribution of time averages of $L^1(m)$ functions 
depends also on the ratio of a measurement time and the time at which system started, i.e., aging distributional behavior 
\cite{Akimoto2013}. Here, we provide another distributional behavior that is in-between the above two distributional limit theorems.

Infinite ergodic theory {has} attracted the interest from not only mathematics but also 
physics community \cite{Gaspard1988,Barkai2003,Akimoto2007,Akimoto2008,Korabel2009,Akimoto2010a,Akimoto2010,Akimoto2012}. 
This is because distributional behaviors of time-averaged 
observables are ubiquitous in phenomena ranging from fluorescence in nano material \cite{Brokmann2003} 
to biological transports \cite{Golding2006,Weigel2011,Tabei2013}. Theoretical studies on distributional behaviors of time-averaged observables have been 
extensively conducted using stochastic models with divergent mean trapping-time distributions such as continuous-time random walks 
\cite{He2008,Miyaguchi2013}, random walk with static disorder \cite{Miyaguchi2011}, and dichotomous processes \cite{Margolin2006}. 
The distribution function of time-averaged observables depends on the type of observation function. In particular, the distribution 
of time-averaged mean square displacement follows the Mittag-Leffler distribution \cite{He2008,Miyaguchi2013}, 
while that of the ratio of occupation time of 
on state in dichotomous processes follows the generalized arc-sine distribution \cite{Margolin2006}.  
Although distributional limit theorems in stochastic processes have been elucidated, 
it will be possible to construct another distributional limit theorem of time-averaged observables by introducing another type 
of the observation function in stochastic models with divergent mean trapping-time distributions. In fact, one of the authors
has shown a novel distributional behavior for time-averaged mean square displacements 
in stored-energy-driven L\'evy flight \cite{Akimoto2013a,Akimoto2014}. 

In this paper, we provide a novel distribution for {time averages of a class of non-$L^1(m)$ functions 
in one-dimensional maps with indifferent fixed points having infinite invariant measures.} The value of the observation function  
at the indifferent fixed point is zero. Because the observation function is non-$L^1(m)$, 
the generalized arc-sine distribution can be applied to those observation functions. However, it only gives a trivial result that 
time averages {converge} to zero. 
Our distributional limit theorem gives a non-trivial broad distribution of normalized time averages. In other words, 
we refine the distribution of normalized time averages of such observation functions by introducing a normalizing 
sequence. The proof is based on a stochastic process induced by a renewal process proposed here, 
which mimics a Birkhoff sum of a non-$L^1(m)$ function.

\section {From dynamical system to stochastic process: partial sums of non-$L^1(m)$ functions}

A dynamical system considered here is a transformation $T:[0,1]\rightarrow [0,1]$ which satisfies the following
conditions for some $c \in (0,1)$: (i) the restrictions $T: (0,c)
\rightarrow (0,1)$ and $T: (c,1) \rightarrow (0,1)$ are $C^2$ and onto,
and have $C^2$-extensions to the respective closed intervals; (ii)
$T'(z)>1$ on $(0, c] \cup [c, 1]$; $T'(0)=1$; (iii) $T(z)-z$ is
regularly varying at zero with index $1+1/\alpha$, $T(z)-z \sim
az^{1+1/\alpha}$ ($\alpha>0$).   {For example, a transformation, 
\begin{equation}
T_\alpha (z)=z
\left(1 + \left(\frac{z}{1+z}
\right)^{\frac{1-\alpha}{\alpha}} -z^{\frac{1-\alpha}{\alpha}}
\right)^{-\frac{\alpha}{1-\alpha}}~({\rm mod}~1),
\label{Thaler_map}
\end{equation}
satisfies the conditions ($a=1$). }
It is known that an invariant measure $m$ of the map is
given by $dm/dz \propto z^{-1/\alpha}$ $(z\rightarrow 0)$ \cite{Thaler1983}.
Thus, the invariant measure cannot be normalized for $\alpha \leq 1$. While this dynamical system has zero-Lyapunov 
exponent, the dynamical instability can be characterized as a sub-exponential instability {\cite{Gaspard1988,Korabel2009,Akimoto2010a}.}
\begin{figure}
\begin{center}
\includegraphics[width=.9\linewidth, angle=0]{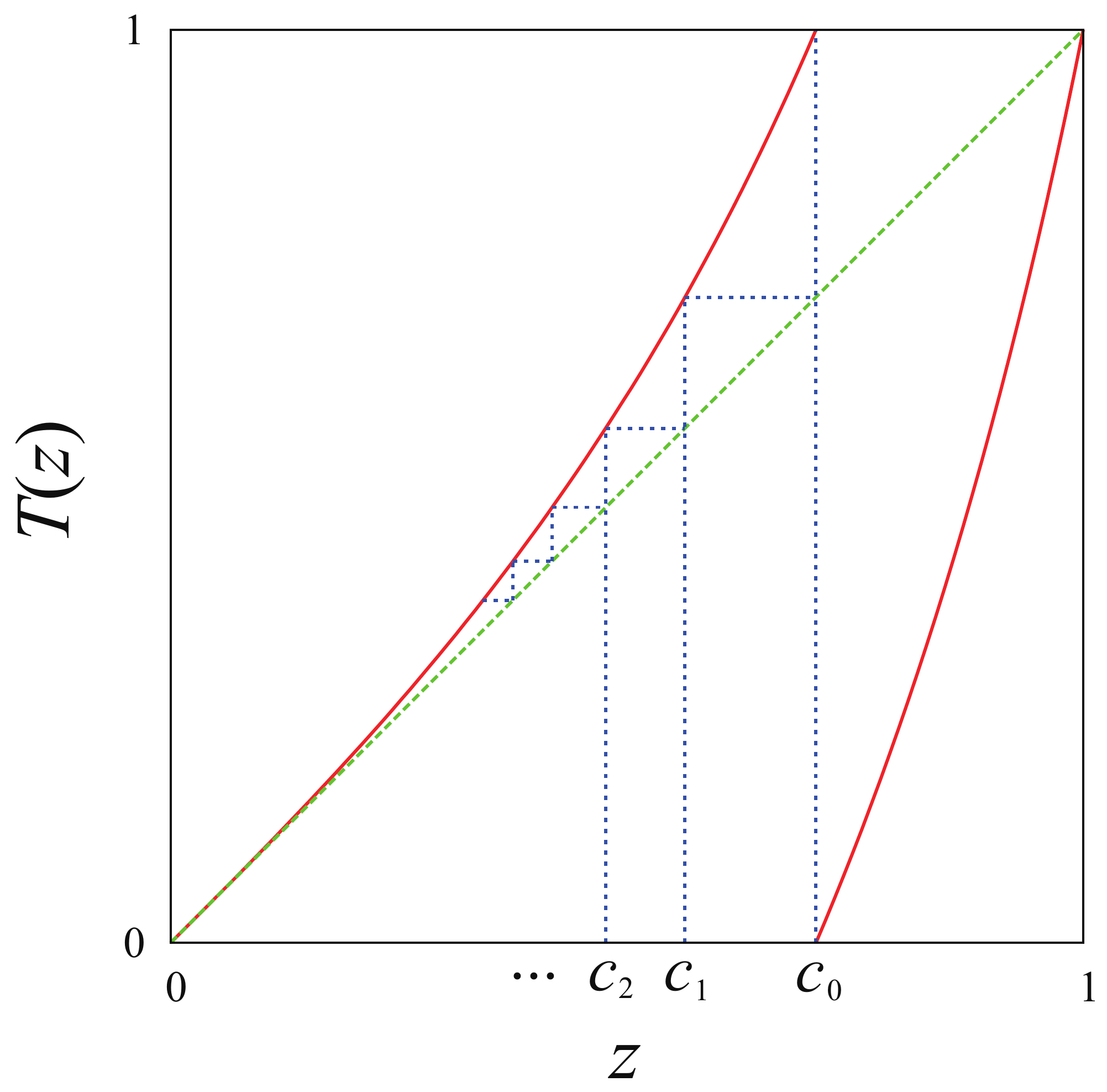}
\end{center}
\caption{Transformation (\ref{Thaler_map}) with $\alpha=0.5$. The origin, $z=0$, is the indifferent fixed point, i.e., $T'(0)=1$. 
The sequence $c_n$  is plotted for $n=0,1,$ and 2. }
\label{map}
\end{figure}

For $z_t \equiv T^t(z_0) \cong 0$, the following ordinary differential equation can be used to
describe the dynamics \cite{Manneville1980,Geisel1984}:
\begin{equation}
\frac{d\tilde{z}_t}{dt} = a \tilde{z}_t^{1+1/\alpha},
\label{ode}
\end{equation}
where we use $\tilde{z}_t$ as the solution of the ordinary differential equation (\ref{ode}) with an 
initial condition $\tilde{z}_0=z_0$. 
The solution is given by
\begin{equation}
  \tilde{z}_t = \tilde{z}_0 \left(1- \frac {t}{\tau}\right)^{-\alpha},
  \label{solution_zt}
\end{equation}
where 
\begin{equation}
\tau = \alpha/(\tilde{z}_0^{1/\alpha} a) 
\label{residence_time}
\end{equation}
 is a characteristic time scale that a trajectory with initial point $\tilde{z}_0$
escapes from $[0,c]$. In fact, a time when $\tilde{z}_t$ becomes unity denoted by $t_c$, i.e., $\tilde{z}_{t_c}=1$, 
is given by $t_c=\tau - \alpha/a$. In what follows, we use a sequence $c_n$ defined by $c_n = T(c_{n+1})$ with $c_n < c$ ($n=1,2, \cdots$) 
and $c_0=c$ {(see Fig.~\ref{map}). 
Trajectory is reinjected to $[0,c]$ from $(c,1]$. Because this dynamical system has a sub-exponential dynamical instability, 
  the reinjection points, $z_0$, can be regarded as a random variable, and it is known that 
 the reinjection points are almost uniformly distributed on $[0,c]$.  Because the distribution of  $\tau$ is determined by that of $z_0$,  } 
  by assuming the probability density function (PDF) of $z_0$
  is uniform on $[0,c]$, we have the PDF of residence times on $[0,c]$:
\begin{equation} 
 w(\tau) \sim \frac {A_c}{|\Gamma(-\alpha)|}
    \tau^{-1-\alpha} \quad {\rm as}~\tau \rightarrow \infty ~(z_0 \to 0),
    \label{trap_dist}
\end{equation}
 where {$A_c$ depends on not only $\alpha$ and $c$} but also details of the map $T(x)$. 
 We note that the mean residence time $\langle \tau \rangle$ diverges when an invariant measure cannot be normalized 
 ($\alpha \leq 1$). Here, we give a rigorous result that a normalized Birkoff's sum can be represented by the trajectory generated by 
 Eq.~(\ref{ode}). 
 
\begin{lemma}
For $t \ll N$, there exists $N >0$ such that  {$z_t < \tilde{z}_t < z_{t+1}$} where 
$z_0=\tilde{z}_0=c_N$. 
\end{lemma}
\begin{proof}
By Eqs.~(\ref{solution_zt}) and (\ref{residence_time}), we have {
\begin{eqnarray}
\tilde{z}_{1} -z_1 &=& c_N \left(1 - \frac{c_N^{1/\alpha} a}{\alpha}\right)^{-\alpha} - c_{N-1}
  >  c_N + a c_{N}^{1+1/\alpha} -c_{N-1}\simeq 0,
  \end{eqnarray}
 and}
\begin{eqnarray}
z_2- \tilde{z}_{1} &{\simeq}& c_{N-1} +a c_{N-1}^{1+1/\alpha} - c_N \left(1 - \frac{c_N^{1/\alpha} a}{\alpha}\right)^{-\alpha}\\
  &=& c_{N-1}- c_N + a (c_{N-1}^{1+1/\alpha} - c_N^{1+1/\alpha}) + o(c_N^{1+1/\alpha}).
  \end{eqnarray}
  Because the sequence $c_n$ is given by $c_n\sim \alpha^\alpha (an)^{-\alpha}$ {for $n\to\infty$} \cite{Thaler1983},  we have 
  \begin{equation}
  c_{N-1}- c_N \sim \left(\frac{\alpha}{aN}\right)^\alpha \left[ \left(1-\frac{1}{N}\right)^{-\alpha}-1\right]
  \sim a\left(\frac{\alpha}{aN}\right)^{\alpha+1} = O(N^{-1-\alpha}),
  \end{equation}
  and 
  \begin{equation}
c_N^{1+1/\alpha} \sim \left(\frac{\alpha}{aN}\right)^{\alpha+1}. 
  \end{equation}
  It follows that {$\tilde{z}_1>z_1$,} $z_2 > \tilde{z}_{1}$ and $z_2 - \tilde{z}_{1}=O(N^{-1-\alpha})$. 
  We assume {$z_{t} < \tilde{z}_{t}<z_{t+1} $} and $z_{t+1} - \tilde{z}_{t}=O(N^{-1-\alpha})$ for $N\to \infty$.
  Then, we have { 
  \begin{eqnarray}
\tilde{z}_{t+1} -z_{t+1} &\simeq&  \tilde{z}_{t}\left(1 - \frac{\tilde{z}_{t}^{1/\alpha} a}{\alpha}\right)^{-\alpha} - z_{t} - a z_{t}^{1+1/\alpha}\\
  &>& \tilde{z}_{t} + a \tilde{z}_{t}^{1+1/\alpha} - z_t - az_t^{1+1/\alpha},
  \end{eqnarray}
  and}
  \begin{eqnarray}
z_{t+2}- \tilde{z}_{t+1} &{\simeq}& z_{t+1} + a z_{t+1}^{1+1/\alpha}
-  \tilde{z}_t \left(1 - \frac{\tilde{z}_t^{1/\alpha} a}{\alpha}\right)^{-\alpha}\\
  &=& z_{t+1} - \tilde{z}_t + a (z_{t+1}^{1+1/\alpha} - \tilde{z}_t^{1+1/\alpha}) + o(N^{-1-\alpha}).
  \end{eqnarray}
 Because we assume {$z_{t} < \tilde{z}_{t} <z_{t+1}$} and $z_{t+1} - \tilde{z}_{t}=O(N^{-1-\alpha})$, we have 
{$\tilde{z}_{t+1} - z_{t+1}>0$,} $z_{t+2}- \tilde{z}_{t+1}>0$ and $z_{t+2} - \tilde{z}_{t+1}=O(N^{-1-\alpha})$. 
 It follows by mathematical induction that there exits $N$ such that {$z_{t}< \tilde{z}_{t} <z_{t+1}$} for $t \ll N$. 
\end{proof}
 
Here, we consider the following bounded continuous observation function, $f(z) \sim Cz^{\frac{1}{\alpha}(1-\gamma)}$ ($z\to 0$), 
which is not an $L^1(m)$ function for $\alpha \leq \gamma <1$;
   we call this type of functions as {\it weak non-$L^1(m)$ functions}.
In particular, we study statistical properties of partial sums of this type of observables,
\begin{equation}
S_t = \sum_{k=0}^{t-1} f(z_k)
\label{partial_sum}
\end{equation}
to elucidate the ergodic properties (Note that $S_t/t$ is the time average).

\begin{lemma}
For $z_0 \in [c_{N+1},c_N)$ and $l (\ll N)$, there exists $N$ such that   
\begin{equation}
\left| \sum_{k=0}^{l} f(z_k) - \int_0^l f(\tilde{z}_t)dt \right| < \Delta I,
\label{lem1}
\end{equation} 
where $\Delta I\equiv f(c_{N-l}) - f(c_{N+1})$.
\end{lemma}

\begin{proof} 
First, we define $I_{\max}$ and $I_{\min}$ as 
\begin{equation}
I_{\max} = \sum_{k=N-l}^N f(c_k)~{\rm and}~
I_{\min} = \sum_{k=N-l}^N f(c_{k+1}),
\end{equation}
and 
\begin{eqnarray}
\Delta I &\equiv& I_{\max} - I_{\min} = f(c_{N-l}) - f(c_{N+1}).
\end{eqnarray}
By Lemma 1, 
\begin{equation}
I_{\min} < \sum_{k=0}^{l} f(z_k) < I_{\max} ~{\rm and}~
I_{\min} < \int_{0}^{l}f(\tilde{z}_t)dt < I_{\max}.
\end{equation}
It follows 
\begin{equation}
\left| \sum_{k=0}^{l} f(z_k) - \int_0^l f(\tilde{z}_t)dt \right| < \Delta I.
\end{equation}
\end{proof}

Here, we decompose the function $f(z)$ into an $L^1(m)$ part and a non-$L^1(m)$ part, where  
an $L^1(m)$ part, $f_\delta^R(z)$, is defined by $f_\delta^R(z)\equiv 0$ on $[0,\delta]$ and $f_\delta^R(z)\equiv f(z)$ on $(\delta,1]$,
and a non-$L^1(m)$ part, $f_\delta^L(z)$, is defined by $f_\delta^L(z)\equiv f(z)$ on $[0,\delta]$ and $f_\delta^L(z)\equiv 0$ on $(\delta,1]$.
By the Aaronson's distributional limit theorem, $\sum_{k=0}^{t-1} f_\delta^R(z_k)/n^\alpha$ converges in distribution for all $\delta>0$ 
because $f_\delta^R(z)$ is an $L^1(m)$ function for all $\delta>0$. 
It follows that for a sequence $a_n$ such that $a_n/n^\alpha \to \infty$ as $n\to \infty$, the normalized time averages, 
$\sum_{k=0}^{t-1} f_\delta^R(z_k)/a_n$, converge 
to zero: $\sum_{k=0}^{t-1} f_\delta^R(z_k)/a_n \to 0$ as $n\to \infty$.

For the dynamical systems defined above, a trajectory is trapped in the interval $[0,\delta]$ for a long time 
and then {escapes} to the other interval $[\delta,1]$ for small $\delta$. Let us consider $k$-th such trapping state. 
We note that the $k$-th trapping time denoted by $\tau_k$ is approximately given by $\alpha/a (z_{0,k}^{-1/\alpha} 
-\delta^{-1/\alpha})$, 
where $z_{0,k}$ is the $k$-th reinjection point.
We will show that a partial sum during the $k$-th trap in $[0, \delta]$, $I(\tau_e, \tau_k) =
\sum_{i=t_{k-1}}^{t_{k-1}+\tau_e} f(z_i)$, can be replaced as {$\int_{0}^{\tau_e} f(\tilde{z}_t)dt$ with $\tilde{z}_0=z_{t_{k-1}}$, where} 
\begin{eqnarray}
\int_{0}^{\tau_e} f(\tilde{z}_t)dt
{\sim} B \tau_k^\gamma   \left[1-\left(1- \frac{\tau_e}{ \tau_k} \right)^{\gamma}\right],
\label{partial_sum_approx}
\end{eqnarray}
{for $\tilde{z}_0 \to 0$,} 
$t_{k-1}=\tau_1 + \cdots + \tau_{k-1}$, $\tau_e \in [0,\tau_k]$ is the elapsed time since the beginning of the trapping, and $B$ is a constant given by
$B=(\alpha/a)^{1-\gamma} C/\gamma$. Because we assume that $z_{0,k}$ is uniformly distributed on {$[0,\delta]$} or equivalently assume 
Eq.~(\ref{trap_dist}),
the PDF of $I(\tau_k)\equiv I(\tau_k,\tau_k)=B\tau_k^\gamma$ is given by
\begin{eqnarray}
l(x) 
\sim \frac{{A_\delta} B^{\frac{\alpha}{\gamma}}}{\gamma |\Gamma(-\alpha)|}  x^{-1-\frac{\alpha}{\gamma}} \quad (x\rightarrow \infty). 
\label{PDF_intesity}
\end{eqnarray}
{We note that the constant {$A_\delta$} depend on $\delta$.}

\begin{lemma}
For $a_t \propto t^\gamma$, the asymptotic behavior of the normalized time average, $S_t/a_t$, is given by 
\begin{equation}
\frac{S_t}{a_t} \sim \frac{1}{a_t} \sum_{k=1}^{N_t} I(\tau_k) + \frac{\int_{0}^{\tau_e} f(\tilde{z}_t)dt}{a_t},
\end{equation}
where {$\tilde{z}_0=z_{t_{N_t}}$}, $N_t$ is the number of reinjections to $[0,\delta]$ until time $t$ and $\tau_k$ is the $k$-th 
trapping time on $[0,\delta]$ and 
$\delta \ll 1$.
\end{lemma}

\begin{proof}
A partial sum is given by
\begin{equation}
S_t = \sum_{k=0}^{t-1} f^L_\delta (z_k)+ \sum_{k=0}^{t-1} f^R_\delta(z_k),
\label{birkhoff_sum}
\end{equation}
where the second term contributes to a Mittag-Leffler distribution but it can be ignored when we consider 
a normalized time averages of weak non-$L^1(m)$ functions, because the order of the normalizing sequence is greater than that of 
the return sequence. In fact, the normalizing sequences for $f_\delta^R$ and $f_\delta^L$ 
are given by $\langle \sum_{k=0}^{t-1} f^R_\delta(z_k) \rangle \propto t^\alpha$ 
and $\langle \sum_{k=0}^{t-1} f^L_\delta (z_k) \rangle \propto t^\gamma$, respectively ($\gamma > \alpha$). 
Therefore, it is sufficient to consider the first term only. 
By Lemma 2, for $\delta \ll 1$ {and $z_i \leq \delta$ ($i=0, \cdots, \tau_e, \cdots, \tau_k$),} there exists a constant $\varepsilon$ such that 
\begin{equation}
\left|\sum_{i=0}^{\tau_k} f_\delta^L(z_i) -  I(\tau_k)\right| < \varepsilon~{\rm and}~
\left|\sum_{i=0}^{\tau_e} f_\delta^L(z_i) -  \int_{0}^{\tau_e} f(\tilde{z}_t)dt \right| < \varepsilon,
\end{equation}
{where the constant $\varepsilon$ does not depend on $\tau_k$ but depend on $\delta$.}
For $a_t = O(t^\gamma)$, we have
\begin{equation}
\frac{1}{a_t}\left|\sum_{k=0}^{t-1}f_\delta^L(z_k) - \sum_{k=1}^{N_t} I(\tau_k) - \int_{0}^{\tau_e} f(\tilde{z}_t)dt \right|
< \frac{\varepsilon (N_t+1)}{a_t}.
\end{equation}
Because $\langle N_t\rangle \propto t^\alpha$ \cite{Akimoto2010}, 
the left-hand-side goes to zero as $t\to\infty$. 
\end{proof}

In the following sections, we will show that there exists a sequence $a_t$ such that the normalized time average,
$S_t /a_t$, converges in distribution:
\begin{equation}
\frac{1}{a_t}\sum_{k=0}^{t-1} f(z_k) \Rightarrow Y_{\alpha,\gamma}
\quad {\rm as}\quad n\rightarrow \infty,
\end{equation}
where the Laplace transform of the random variable $Y_{\alpha,\gamma}$ is given by Eq. (\ref{Y_alpha_gamma}).
We note that the sequence $a_t$ is given by $a_t \equiv \langle \sum_{k=0}^{t-1} f(z_k) \rangle \propto t^\gamma$, which is not 
the so-called return sequence in infinite ergodic theory \cite{Aaronson1997}. 
 In particular, the order of the return sequence is given by $t^\alpha$, which is smaller than that of $a_t$, i.e., $t^\alpha/ a_t \to 0$ as 
 $t \to \infty$. 

\begin{figure}
\begin{center}
\includegraphics[width=.9\linewidth, angle=0]{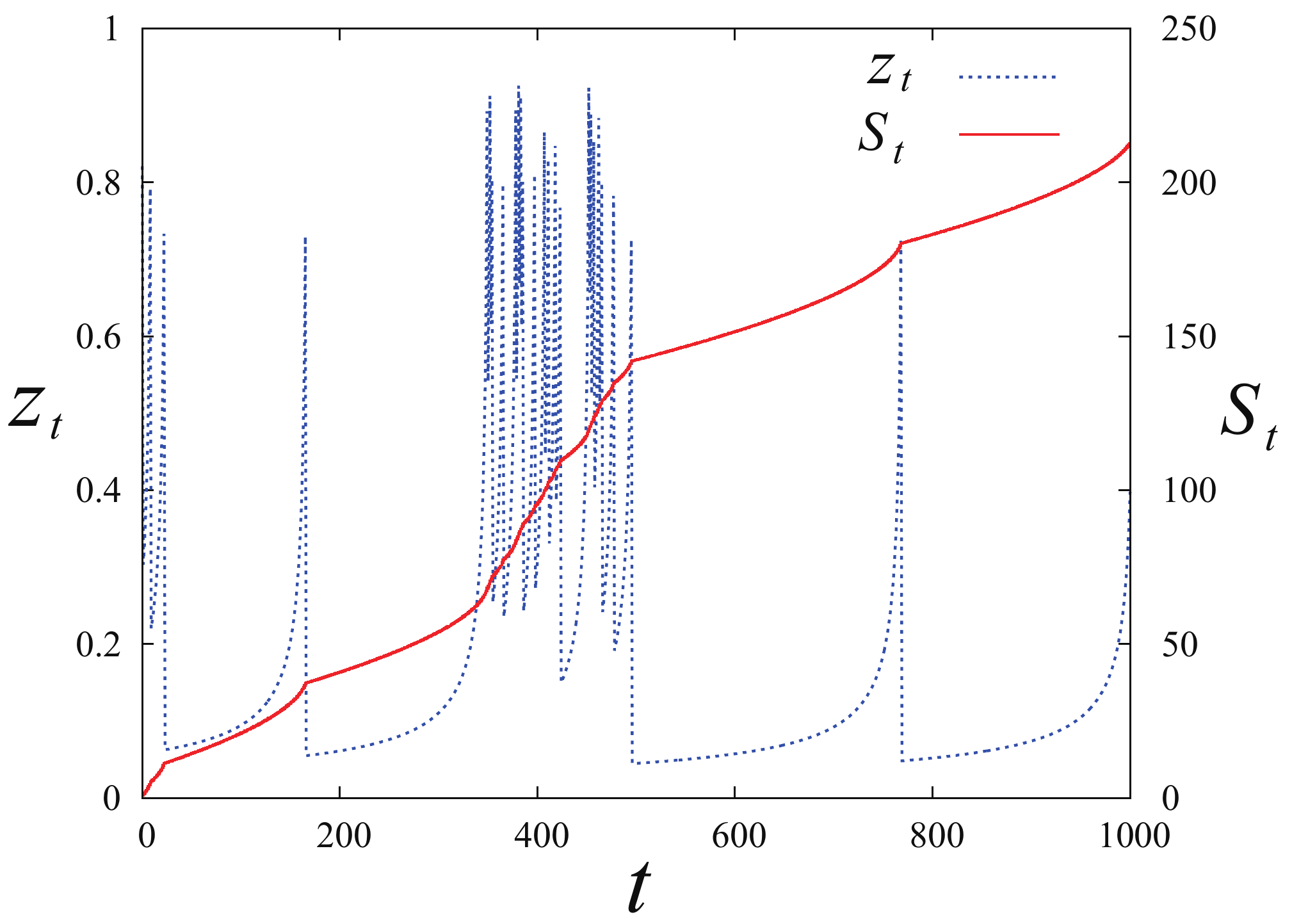}
\end{center}
\caption{Continuous accumulation process generated by the map (\ref{Thaler_map}) with $\alpha=0.5$ and 
$f(z)=z^{0.6}$. 
The solid line represents a partial sum $S_t$, which corresponds to $X_t$ in the continuous accumulation process, 
while the dashed line represents the trajectory. }
\label{GRP}
\end{figure}

\section{Continuous accumulation process}

To analyze the partial sum [Eq.~(\ref{partial_sum})], we generalize a renewal process.
Renewal process is a point process where the time intervals between point events 
are independent and identically distributed (i.i.d) random variables \cite{Cox}. Because residence times near the indifferent fixed 
point in intermittent maps are considered to be almost i.i.d. random variables, one can apply renewal processes to study 
dynamical systems \cite{Akimoto2010b}.

  Here, we consider a {\it cumulative process} by introducing an intensity of each renewal event \cite{Cox}, where intensity 
  is correlated with the time interval between successive renewals. 
  This process can be characterized by the total intensity $X_t$ until time $t$, whereas renewal processes are characterized by 
  the number of renewals in the time interval $[0,t]$, denoted by $N_t$. 
  Let $\tau_1, \ldots, \tau_k$ be the time intervals between successive renewals, which are i.i.d. random variables with PDF $w(\tau)$. 
  We assume that the $k$-th intensity is determined by  the $k$-th interevent time (IET) $\tau_k$ 
  as $I(\tau_k)\equiv B\tau^\gamma_k$, where $\gamma \in [0,1)$ and 
$B>0$. Thus, the longer the IET between renewals becomes, the larger the intensity is. 
Furthermore, we propose a {\it continuous accumulation process}  induced by a renewal process to consider a Birkhoff sum. 
In the continuous accumulation process, the intensity is gradually accumulated  according to a
function $I(\tau_e, \tau_k)$ in between the two successive renewals, where $\tau_e \in [0,\tau_k]$ is the
elapsed time after the $(k-1)$-th renewal {(see Fig.~\ref{GRP})}.  Here, we use the following intensity function:
\begin{equation}
  I(\tau_e, \tau_k)=
  I(\tau_k) \left[ 1 - \left(1-\frac{\tau_e}{\tau_k}\right)^\gamma\right]. 
\label{intensity_func}
\end{equation}
This intensity function $I(\tau_e, \tau)$ mimics an increase of the Birkhoff sum 
in dynamical systems [see Eq.~(\ref{partial_sum_approx})]. As shown in the previous section,
  the following stochastic variable $X_t$ {(the integrated intensity up to time $t$), 
\begin{equation}
X_t = \sum_{k=1}^{N_t} I(\tau_k) + I(t-t_{N_t}, \tau_{N_t +1}),
\end{equation}
where $t_k=\tau_1+\ldots \tau_k$,  is related to the Birkhoff sum of the non-integrable function.
The case in which $\gamma=0$ and $B=1$ 
is exactly} equivalent to the usual renewal process, because  $X_t =N_t$.
  
Here, we consider the case that the mean interevent times of renewals diverges $(\alpha\leq 1)$. In particular, 
we use Eq.~(\ref{trap_dist}) as the PDF of IETs. Thus, the survival probability $W(\tau)$ is given by  
\begin{equation}
W(\tau) \equiv 1 - \int_0^\tau w(\tau')d\tau' \sim \frac{A}{\Gamma(1-\alpha)} \tau^{-\alpha}\quad (\tau \rightarrow \infty),
\end{equation}
and the PDF of $I(\tau_k)$, denoted by $l(x)$,  is given by Eq.~(\ref{PDF_intesity}). Because 
 the mean intensity $\langle I \rangle$ diverges  for $\alpha \leq \gamma$, the renewal theory cannot be 
 straightforwardly applied.

\section{Theory of a continuous accumulation process}

\subsection{Generalized renewal equation}
Distribution of $X_t$ can be derived by a generalized renewal equation, which is 
  similar to a generalized master equation for the continuous-time random walk (CTRW) \cite{Shlesinger1982}. 
  First,  we define a joint PDF of the IET $\tau$ and the
  intensity increment $x$ as $\psi (x,\tau) = w(\tau) \delta(x-I(\tau))$, and we also use 
 $\Psi(x,\tau_e; \tau) = \delta(x- I(\tau_e, \tau))\theta(\tau-\tau_e)$, where $\theta(x) = 0$ for $x<0$ and 
 1 otherwise. 
Let $Q(x,t)$ be the PDF of $X_t$ at time $t$ when a renewal occurs, then we have
\begin{eqnarray}
Q(x,t) &=& \int_0^\infty dx' \psi(x',t) \delta (x-x') \nonumber\\
&&
+  \int_0^x dx' \int_0^t dt' \psi(x',t')  Q(x-x', t-t') .
\end{eqnarray}
The conditional PDF of $X_t$  at time $t$ on the condition of $\tau_{N_t+1}=\tau$, denoted by 
$P(x,t;\tau)$ is given by 
\begin{eqnarray}
P(x,t;\tau) &=& \int_0^x dx' \int_0^t dt' \Psi(x',t'; \tau) Q(x-x',t-t')  
+ \Psi(x,t; \tau).
\end{eqnarray}
It follows that the PDF of $X_t$  at time $t$ reads
\begin{equation}
P(x,t)= \int_0^\infty w(\tau) P(x,t;\tau)d\tau.
\end{equation} 
Here, we assume that a renewal occurs at time $t=0$, i.e., ordinary renewal process \cite{Cox}. 
Using the double Laplace transform with respect to time ($t\rightarrow s$) and $X_t$ ($x\rightarrow k)$, 
defined by
\begin{equation}
{\hat{P}}(k,s) \equiv \int_0^\infty dt \int_0^\infty dx e^{-st -kx} P(x,t), 
\end{equation}
we have 
\begin{eqnarray}
{\hat{P}}(k,s) 
= \int_0^\infty \frac{w(\tau)\hat{\Psi}(k,s;\tau)}{1-{\hat{\psi}}(k,s)} d\tau, 
\label{renewal_eq}
\end{eqnarray}
where 
\begin{eqnarray}
{\hat{\psi}} (k,s) \equiv 
\int_0^\infty d\tau \int_0^\infty dx e^{-s\tau -kx} \psi(x,\tau) 
=\int_0^\infty e^{-s\tau} e^{-kI(\tau)} w (\tau) d\tau,
\end{eqnarray}
and
\begin{eqnarray}
\hat{\Psi} (k,s;\tau) \equiv 
\int_0^\infty dt \int_0^\infty dx e^{-st -kx} \Psi(x,t;\tau) 
=
\int_0^\tau e^{-st -k I(t, \tau)} dt.
\end{eqnarray}
In what follows, we use the asymptotic behaviors of the Laplace transforms of $w(\tau)$ and $W(\tau)$, i.e., 
$1- \hat{w}(s) \sim A s^\alpha$ and $\hat{W}(s) \sim A s^{\alpha-1}$ for $s\rightarrow 0$.


\subsection{Moments of $X_t$}
\subsubsection{First moment}
The Laplace transform of $\langle X_t \rangle$, denoted by $\langle X_s \rangle$, is given by
$\langle X_s \rangle = \left. -\frac{\partial {\hat{P}}(k,s)}{\partial k}\right|_{k=0}$. 
As shown in the Appendix~A, the leading order of $\langle X_s \rangle$ is given by 
\begin{eqnarray}
\langle X_s \rangle 
\sim
\left\{
\begin{array}{ll}
{\displaystyle  \frac{ B M_1(\alpha,\gamma)}{|\Gamma(-\alpha)|} \frac{1}{s^{1+\gamma}}}, \quad &(\gamma > \alpha)\\[.5cm]
{\displaystyle \frac{B}{|\Gamma(-\alpha)|} \frac{1}{s^{1+\alpha}}} \log \left(\frac{1}{s}\right), &(\gamma=\alpha)\\[.5cm]
{\displaystyle \frac{\langle I \rangle}{A} \frac{1}{s^{1+\alpha}}}, &(\gamma<\alpha)
\end{array}
\right.
\end{eqnarray}
where $M_1(\alpha,\gamma)=\Gamma(\gamma-\alpha)\left[ 1+\frac{\gamma(\gamma-\alpha)}{1+\alpha-\gamma}\right]$. 
The inverse Laplace transform reads
\begin{equation}
\langle X_t \rangle \sim 
\left\{
\begin{array}{ll}
{\displaystyle \frac{B M_1(\alpha, \gamma)}{|\Gamma(-\alpha)| \Gamma(1+\gamma)}  t^{\gamma}}, \quad &(\gamma > \alpha)\\[.5cm]
{\displaystyle \frac{B}{|\Gamma(-\alpha)| \Gamma(1+\alpha)} t^\alpha \log t}, &(\gamma=\alpha)\\[.5cm]
{\displaystyle \frac{\langle I \rangle}{A\Gamma(1+\alpha)} t^\alpha.} &(\gamma<\alpha)
\end{array}
\right.
\label{1st_moment}
\end{equation}
{We note that the asymptotic behavior of $\langle X_t \rangle$ is determined by $B,\alpha,$ and $\gamma$. In other 
words, it does not depend on $A$.}

\subsubsection{Second moment}
The Laplace transform for the second moment of $X_t$, denoted by $\langle X^{2}_s \rangle$ is given by
$\langle X^{2}_s \rangle = \left. \frac{\partial^{2} {\hat{P}}(k,s)}{\partial k^{2}} \right|_{k=0}$. 
As shown in  the Appendix~B, the leading order of $\langle X_s^2 \rangle$ is given by 
\begin{eqnarray}
\langle X^2_s \rangle 
&\sim& \left\{
\begin{array}{ll}
{\displaystyle  \frac{B^2 M_2(\alpha,\gamma)}{|\Gamma(-\alpha)|} 
 \frac{1}{s^{1+2\gamma}}}, \quad &(\gamma > \alpha)\\[.5cm]
{\displaystyle \frac{2B^2}{|\Gamma(-\alpha)|^n} \frac{1}{s^{1+2\alpha}}} \log^2 \left(\frac{1}{s}\right), &(\gamma=\alpha)\\[.5cm]
{\displaystyle \frac{2\langle I \rangle^2}{A^2} \frac{1}{s^{1+2\alpha}}}, &(\gamma<\alpha)
\end{array}
\right.
\end{eqnarray}
where $M_2(\alpha,\gamma)=\left\{1+\frac{\gamma^2 (2\gamma-\alpha)  }{2+\alpha  -\gamma}\right\} \Gamma(2\gamma -\alpha)
+ \left\{ 1 + \frac{\gamma (\gamma-\alpha) }{1+\alpha  -\gamma}
\right\} \frac{2 \Gamma(\gamma-\alpha)^2}{|\Gamma(-\alpha)|}$.
The inverse Laplace transform reads
\begin{eqnarray}
\langle X^2_t \rangle 
&\sim& \left\{
\begin{array}{ll}
{\displaystyle  \frac{B^2 M_2(\alpha,\gamma)}{|\Gamma(-\alpha)|\Gamma(1+2\gamma)} 
  t^{2\gamma}}, \quad &(\gamma > \alpha)\\[.5cm]
{\displaystyle \frac{2B^2}{|\Gamma(-\alpha)|^n} t^{2\alpha} (\log t)^2}, &(\gamma=\alpha)\\[.5cm]
{\displaystyle \frac{2\langle I \rangle^2}{A^2} t^{2\alpha}.} &(\gamma<\alpha)
\end{array}
\right.
\label{2nd_moment}
\end{eqnarray}

\subsubsection{$n$th moment}
\begin{figure}
\includegraphics[width=.55\linewidth, angle=0]{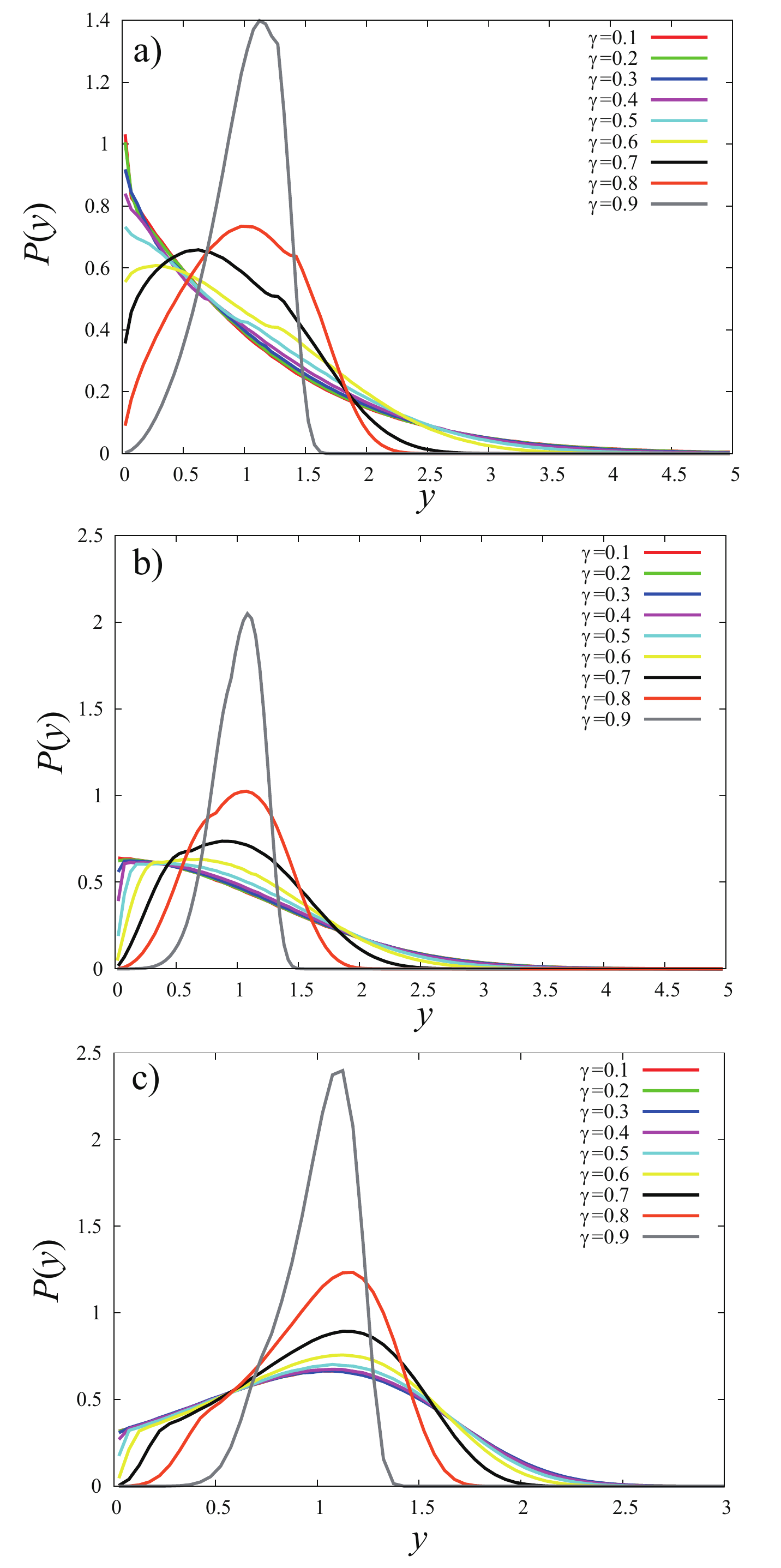}
\caption{Probability density functions of $y=X_t/\langle X_t \rangle$ for (a) $\alpha=0.25$, (b) $\alpha=0.5$ and (c) $\alpha=0.75$. The PDFs 
are obtained by numerical simulations of {continuous accumulation} processes. Total simulation time $t$ is $10^7, 10^6$ and $10^5$ 
for $\alpha=0.25, 0.5$ and 0.75, respectively.
PDFs crucially depend on $\gamma$ for $\gamma > \alpha$. Because of finite simulation time,
 PDFs for $\gamma < \alpha$ slightly depend on $\gamma$. In the numerical simulations, we used the PDF of $\tau$ as 
 $w(\tau)=\alpha \tau^{-1-\alpha}$ ($\tau\geq 1$) and $B=1$. }
\label{limit_dist_accumulation}
\end{figure}

As shown in the Appendix~C, the leading order of the Laplace transform of $\langle X^{n}_t\rangle$ {with $n>0$} for $s\rightarrow 0$ is given by
\begin{eqnarray}
\langle X^{n}_s \rangle &=& \hat{P}^{(n)}(0,s) \nonumber\\
&\sim&
\left\{
\begin{array}{ll}
{\displaystyle  \frac{B^{n} M_n(\alpha, \gamma)}{|\Gamma(-\alpha)|} \frac{1}{s^{1+n\gamma}} \quad (\gamma>\alpha) }\\[.5cm]
{\displaystyle   \frac{(-1)^{n}n![1- \hat{\psi}(s)]}{s [1 - {\hat{\psi}}(0,s)]^{n+1}} \{\hat{\psi}'(0,s)\}^n }\quad (\gamma \leq \alpha) 
\end{array}
\right.
\\
&\sim& \left\{
\begin{array}{ll}
{\displaystyle \frac{B^nM_n(\alpha, \gamma)}{|\Gamma(-\alpha)|} \frac{1}{s^{1+n\gamma}}}, \quad &(\gamma > \alpha) \\[.5cm]
{\displaystyle \frac{ n!B^{n} }{|\Gamma(-\alpha)|^n} \frac{1}{s^{1+n\alpha}} \left\{\log \left(\frac{1}{s}\right) \right\}^n}, &(\gamma=\alpha)
\\[.5cm]
{\displaystyle \frac{ n! \langle I \rangle^n}{A^n} \frac{1}{s^{1+n\alpha}} }, &(\gamma<\alpha).
\end{array}
\right.
\end{eqnarray}
where $M_n(\alpha,\gamma)$ is given by (S14). 
The inverse Laplace transform reads
\begin{equation}
\langle X^{n}_t \rangle \sim  \left\{
\begin{array}{ll}
{\displaystyle 
\frac{ B^nM_n(\alpha,\gamma)}{|\Gamma(-\alpha)|\Gamma(1+n\gamma)}  
t^{n\gamma}} &(\gamma > \alpha) \\[.5cm]
{\displaystyle \frac{ n!B^{n} }{|\Gamma(-\alpha)|^n \Gamma(1+n\alpha)} t^{n\alpha} (\log t )^n} &(\gamma=\alpha)
\\[.5cm]
{\displaystyle \frac{ n!\langle I \rangle^n}{A^n \Gamma(1+n\alpha)} t^{n\alpha}} &(\gamma<\alpha).
\end{array}
\right.
\label{moments_Xt}
\end{equation}
It follows that $X_t/\langle X_t\rangle$ converges in distribution to $Y_{\alpha,\gamma}$, where 
\begin{equation}
\langle e^{zY_{\alpha,\gamma}} \rangle = \left\{
\begin{array}{ll}
{\displaystyle \sum_{k=0}^{\infty} 
\frac{{M_k}(\alpha,\gamma) z^k}{ k!M_1(\alpha,\gamma)^k \Gamma(1+k\gamma)}}
\quad &(\gamma > \alpha)\\[.7cm]
{\displaystyle \sum_{k=0}^{\infty}  \frac{\Gamma(1+\alpha)^kz^k}{\Gamma(1+k\alpha)}}
\quad &(\gamma \leq \alpha),
\end{array}
\right.
\label{Y_alpha_gamma}
\end{equation}
{and $M_0(\alpha,\gamma)=1$.}
We note that the distribution of  the normalized random variable $X_t/\langle X_t\rangle$ does not depend 
on $\gamma$ for $\gamma \leq \alpha$ (in this case the observation function is $L^1(m)$ in the dynamical system) 
and the distribution is called the Mittag-Leffler distribution of order $\alpha$,
whereas the distribution of a scaled sum $X_t/ \langle X_t \rangle$ for $\gamma>\alpha$ 
converges to a time-independent non-trivial distribution, which is not the Mittag-Leffler distribution. 
Figure~\ref{limit_dist_accumulation} shows the PDFs of $X_t/\langle X_t\rangle$ for different $\alpha$ and $\gamma$. 

\section{Distribution of time averages of weak non-$L^1(m)$ functions}

In the previous section, we have shown that the normalized random variable $X_t/\langle X_t \rangle$ converges to 
$Y_{\alpha,\gamma}$ in distribution.
 Because $S_t = \sum_{k=0}^{t-1} f(z_k)$ can be represented by $X_t$ for $\gamma >\alpha$ (Lemma 3),  we have the following 
  proposition for the distribution of time average of a weak non-$L^1(m)$ function. 
 
 \begin{proposition}
For a transformation satisfying the conditions (i), (ii), and (iii), and a bounded continuous observation function 
$f(z) \sim C z^{\frac{1}{\alpha}(1-\gamma)}$ $(z\to 0)$ with $\alpha \leq 1$ and $\alpha < \gamma < 1$,  
there exists sequence $a_n$ such that the normalized time average, $S_n/a_n$, converges in 
distribution:
\begin{equation}
\frac{1}{a_n}\sum_{k=0}^{n-1} f(z_k) \Rightarrow Y_{\alpha,\gamma}
\quad {\rm as}\quad n\rightarrow \infty,
\end{equation}
where the Laplace transform of the random variable $Y_{\alpha,\gamma}$ is given by Eq.~(\ref{Y_alpha_gamma}).
\end{proposition}
\begin{remark}
The sequence $a_n$ is given by $a_n = \langle S_n \rangle \propto n^\gamma$ for 
$\gamma > \alpha$, which is not the return sequence, where $\langle \cdot \rangle$ means an average with 
respect to the initial point $z_0$. 
\end{remark}

\begin{proof}
By Lemma 3, the distribution of time averages of $f(z)$ in the dynamical system considered here can be regarded as 
that in a continuous accumulation process. 
The result in section 4 implies the proposition.
\end{proof}

\begin{figure}
\includegraphics[width=1.\linewidth, angle=0]{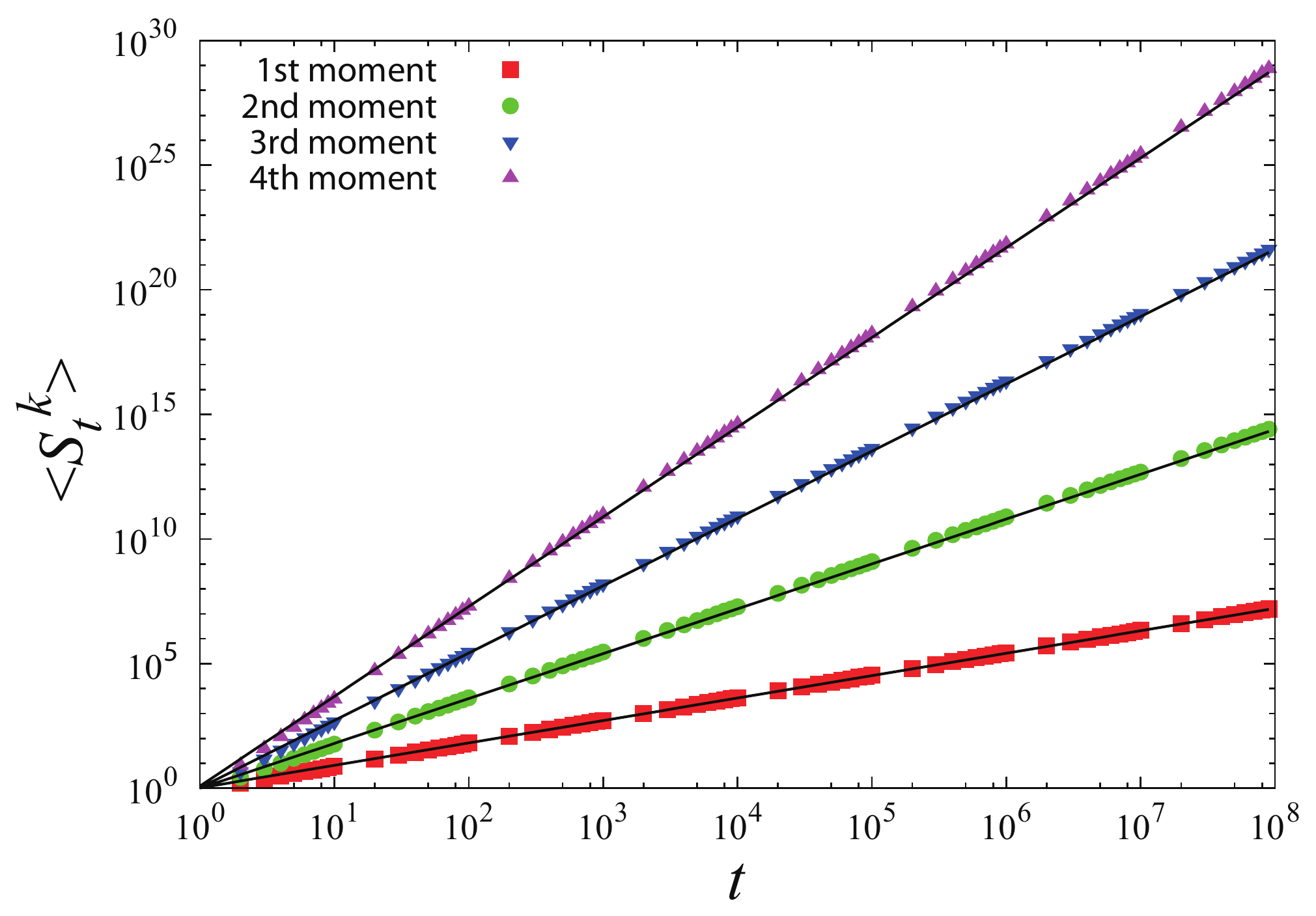}
\caption{Moments of $S_t$ in the map (\ref{Thaler_map}), where $\alpha=0.5$, $\gamma=0.9$, and $f(z)=z^{\frac{1-\gamma}{\alpha}}$. 
Symbols are the results of numerical simulations. Solid lines are the theoretical ones
 (\ref{1st_moment}), (\ref{2nd_moment}), and (\ref{moments_Xt}). The asymptotic behaviors are well described by the theory without fitting. 
 }
\label{moments_St}
\end{figure}
\begin{figure}
\includegraphics[width=.9\linewidth, angle=0]{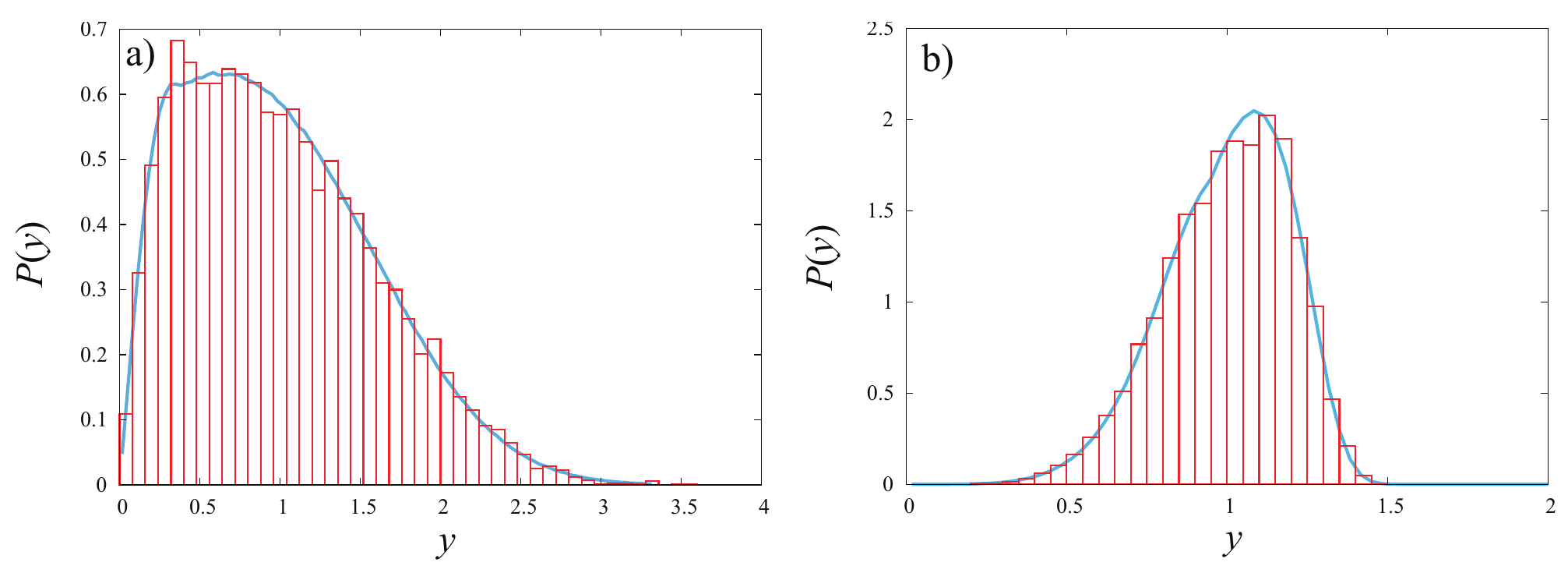}
\caption{Probability density functions of $S_t/\langle S_t \rangle$ with $f(z)=z^{\frac{1-\gamma}{\alpha}}$ 
in the map (\ref{Thaler_map}) with $\alpha=0.5$ for $\gamma=0.6$ and 0.9. 
Histograms are the results of numerical simulations with (a) $\gamma=0.6$ and (b) $\gamma=0.9$. The solid lines are the PDFs 
 obtained by numerical simulation of {continuous accumulation} processes. Total simulation time is $t=10^7$. 
Histograms are in good agreement with the PDFs obtained in the corresponding {continuous accumulation} processes.}
\label{limit_dist}
\end{figure}

To demonstrate our proposition, 
we use the map $T_{\alpha}: [0,1]\rightarrow [0,1]$ with $\alpha \leq 1$ \cite{Thaler2000} defined by 
Eq.~(\ref{Thaler_map})
The asymptotic behavior of $T(x)$ for $x\rightarrow 0$ is given by 
$T_\alpha (x) -x \sim x^{1+1/\alpha}$. Thus, $a=1$. 
The invariant density $\rho_\alpha (x)$ of this map is exactly known as \cite{Thaler2000}
\begin{equation}
\rho_\alpha (x)= \frac{C_\alpha}{x^{\frac{1}{\alpha}}} + \frac{C_\alpha}{(1+x)^{\frac{1}{\alpha}}},
\end{equation}
where $C_\alpha$ is a multiplicative constant. {By numerical simulations, we have confirmed that 
the asymptotic behaviors of moments of $S_t$ 
are well described by the theory (\ref{moments_Xt}) as shown in Fig.~\ref{moments_St}. }
Figure~\ref{limit_dist} shows that 
PDF of $S_t/\langle S_t \rangle$ is in good agreement with the PDF of $X_t/\langle X_t \rangle$ in the 
corresponding continuous accumulation process. 

{
Because the ensemble average of $f(z)$ with respect to an infinite measure diverges, one cannot obtain 
a relation between the time average and the ensemble average with respect to the infinite measure. However, we have shown that 
$\langle S_t/t^\gamma\rangle$ converges to a constant, and the constant is determined by $\alpha, \gamma$, 
and $B$. Because these constants are determined by the asymptotic behaviors of $T(z)$ and $f(z)$,
the normalizing sequence, $a_n$, in the proposition can be determined by the asymptotic behaviors of $T(z)$ and $f(z)$. 
In other words, the sequence does not depend on the details of a map $T(z)$ and $f(z)$ except for a small $z$ behavior. 
We have numerically confirmed them (not shown).
}

\section{Conclusion}
\begin{figure}
\includegraphics[width=.9\linewidth, angle=0]{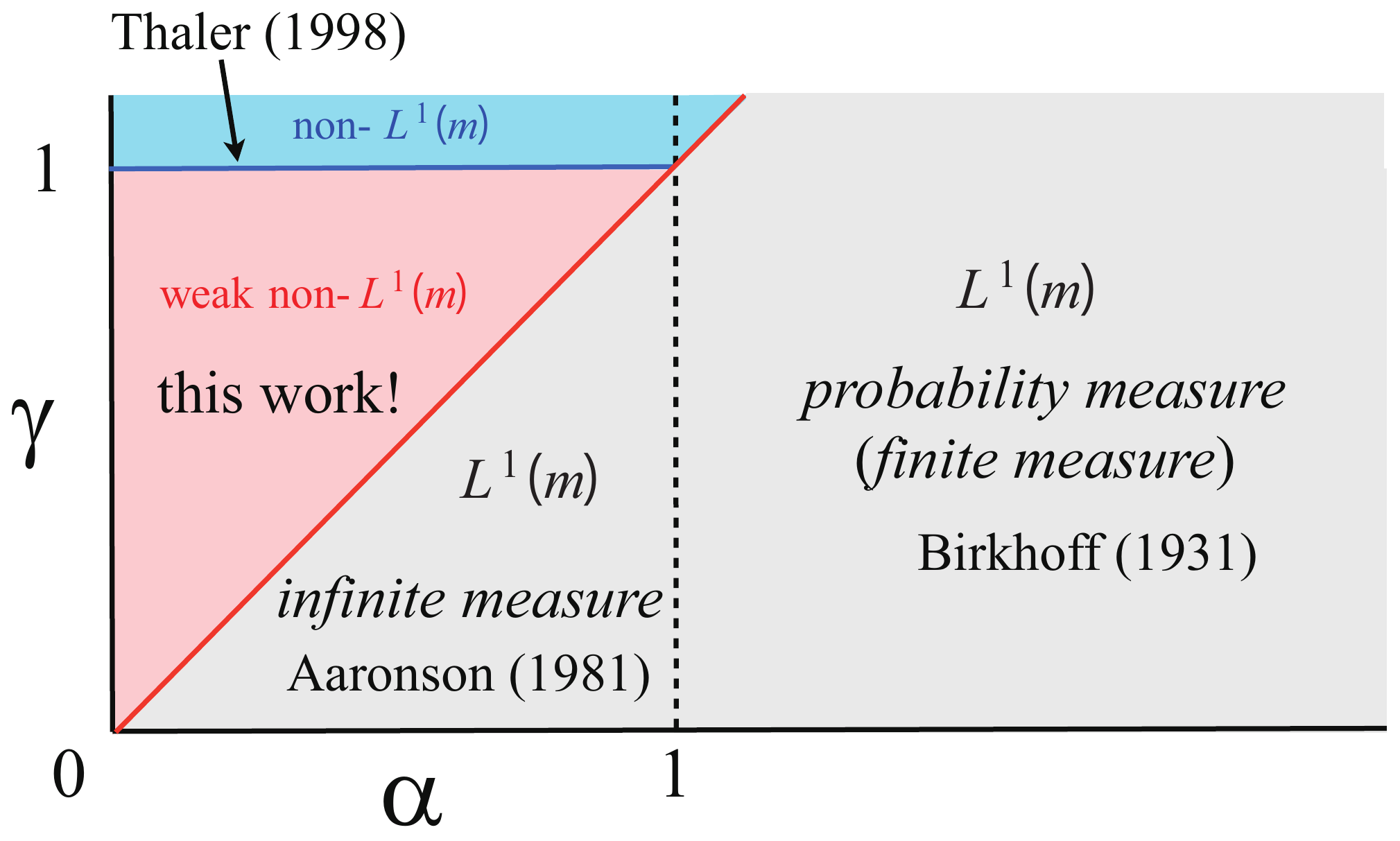}
\caption{Summary of this work. A class of observation functions can be classified by the exponents $\alpha$ and $\gamma$. 
We note that $\alpha \leq 1$ implies an infinite invariant measure. 
The exponent $\gamma$ of the observation function considered here satisfies $\alpha \leq \gamma <1$.}
\label{summary}
\end{figure}

For one-dimensional intermittent maps with infinite invariant measures, we have shown 
a novel distributional behavior for time averages of weak non-$L^1(m)$ functions. The distribution
refines the generalized arc-sine distribution of time average for weak non-$L^1(m)$ functions because the normalizing 
sequence is not $n$ but is proportional to $n^\gamma$ ($\gamma<1$). Therefore, the distribution is not the generalized 
arc-sine distribution nor the Mittag-Leffler distribution. In other words, we have made an important first 
step for a foundation of the third 
distributional limit theorem in infinite ergodic theory. {Recently, distributional behaviors in 
intermittent maps with more than two indifferent fixed points has been studied \cite{Shinkai2007,Korabel2012,Korabel2013}. 
This kind of extension will be interesting for a future work.} 
The proof of our proposition is based on the theory of the continuous accumulation process proposed here. 
Our result is summarized in Fig.~\ref{summary}. This novel distributional limit theorem is related to a distributional behavior 
of time-averaged diffusion coefficients in a model of anomalous diffusion like stored-energy-driven L\'evy flight \cite{Akimoto2013a}. 

\section*{acknowledgement}　
This work was inspired by the conference of ``Weak Chaos, Infinite Ergodic Theory, and Anomalous Dynamics." 
We are indebted to T. Miyaguchi and H. Takahashi for his helpful comments.
This work was partially supported by Grant-in-Aid for Young Scientists (B) (Grant No. 26800204).

\appendix

\section{First moment}

The Laplace transform of $\langle X_t \rangle$ is given by
\begin{align}
\langle X_s \rangle &= \left. -\frac{\partial {\hat{P}}(k,s)}{\partial k}\right|_{k=0}\nonumber\\
&= -\frac{\hat{\psi}'(0,s) \int_0^\infty W(\tau) e^{-s\tau} d\tau}{ [1 - {\hat{\psi}}(0,s)]^2} +
\frac{\int_0^\infty w(\tau) [\int_0^\tau I(t,\tau)e^{-st}dt] d\tau}{ 1 - {\hat{\psi}}(0,s)} \nonumber\\
&=-\frac{\hat{\psi}'(0,s)}{ s[1 - \hat{w}(s)]}
+\frac{\int_0^\infty dt \left[ \int_t^\infty d\tau w(\tau) I(t,\tau)\right] e^{-st}}{ 1 - {\hat{w}}(s)}. \tag{S1}
\end{align}
Because the asymptotic behavior of $w(\tau) I(t,\tau)$ is given by 
$w(\tau) I(t,\tau)\sim AB\gamma t\tau^{-2+\gamma-\alpha}/|\Gamma(-\alpha)|$ for $t/\tau \ll 1$, 
the integral in the second term can be calculated as follows:
\begin{align}
\int_0^\infty dt \left[ \int_t^\infty d\tau w(\tau) I(t,\tau)\right] e^{-st} &\sim \frac{AB\gamma}{|\Gamma(-\alpha)|} 
 \int_0^\infty  t\left[\int_t^\infty \tau^{-2-\alpha+\gamma}d\tau \right] e^{-st} {dt}\nonumber\\
 &= \frac{AB\gamma }{|\Gamma(-\alpha)| (1+\alpha  -\gamma)} \int_0^\infty t^{\gamma -\alpha}e^{-st}dt\nonumber\\
 &\sim \frac{AB\gamma \Gamma(\gamma -\alpha +1) }{|\Gamma(-\alpha)| (1+\alpha  -\gamma)} \frac{1}{s^{\gamma-\alpha+1}}. 
 \tag{S2}
\end{align}
Using the asymptotic behavior of $\hat{\psi}'(0,s)$, 
\begin{align}
\hat{\psi}'(0,s) \sim - \frac{AB \Gamma(\gamma-\alpha)}{|\Gamma(-\alpha)|}\frac{1}{s^{\gamma-\alpha}}\quad (\gamma >\alpha),
\tag{S3}
\end{align}  
we have
\begin{align}
\langle X_s \rangle &= \frac{B \Gamma (\gamma -\alpha)}{|\Gamma(-\alpha)|} \frac{1}{s^{1+\gamma}}
+\frac{B\gamma (\gamma -\alpha) \Gamma(\gamma -\alpha) }{|\Gamma(-\alpha)| (1+\alpha  -\gamma)} \frac{1}{s^{1+\gamma}} 
\nonumber\\
&= \frac{B \Gamma (\gamma -\alpha)}{|\Gamma(-\alpha)|} \left[ 1+\frac{\gamma(\gamma-\alpha)}{1+\alpha-\gamma}\right]
\frac{1}{s^{1+\gamma}}.
\tag{S4}
\end{align}

\section{Second moment}

The Laplace transform for the second moment of $X_t$ is given by
\begin{align}
\langle X^{2}_s \rangle &= \left. \frac{\partial^{2} {\hat{P}}(k,s)}{\partial k^{2}} \right|_{k=0}\nonumber\\
&= \frac{\int_0^\infty w(\tau) \int_0^\tau I(t,\tau)^2 e^{-st} dt d\tau}{1-\hat{w}(s)} 
- \frac{2\hat{\psi}'(0,s) \int_0^\infty w(\tau) \int_0^\tau  I(t,\tau) e^{-st}dt d\tau }{[1-\hat{w}(s)]^2} 
+ \frac{2 \{\hat{\psi}'(0,s)\}^2}{s[1-\hat{w}(s)]^2} + \frac{\hat{\psi}''(0,s)}{s[1-\hat{w}(s)]}\nonumber\\
&= \frac{\int_0^\infty dt \left[ \int_t^\infty d\tau w(\tau) I(t,\tau)^2\right]e^{-st}}{1-\hat{w}(s)} 
- \frac{2\hat{\psi}'(0,s) \int_0^\infty dt \left[ \int_t^\infty d\tau w(\tau) I(t,\tau)\right]e^{-st} }{[1-\hat{w}(s)]^2} \nonumber\\
&+ \frac{2 \{\hat{\psi}'(0,s)\}^2}{s[1-\hat{w}(s)]^2} + \frac{\hat{\psi}''(0,s)}{s[1-\hat{w}(s)]}.
\tag{S5}
\end{align}
Because the asymptotic behavior of $w(\tau) I(t,\tau)^2$ is given by 
$w(\tau) I(t,\tau)^2\sim A(B\gamma t)^2\tau^{-3+2\gamma-\alpha}/|\Gamma(-\alpha)|$ for 
{$t/\tau \ll 1$}, 
the integral in the first term can be calculated as follows:
\begin{align}
\int_0^\infty dt \left[ \int_t^\infty d\tau w(\tau) I(t,\tau)^2 \right] e^{-st} &\sim \frac{A(B\gamma)^2}{|\Gamma(-\alpha)|} 
 \int_0^\infty  t^2\left[\int_t^\infty \tau^{-3-\alpha+2\gamma}d\tau \right] e^{-st} dt\nonumber\\
 &= \frac{A(B\gamma)^2 }{|\Gamma(-\alpha)| (2+\alpha  -2\gamma)} \int_0^\infty t^{2\gamma -\alpha}e^{-st}dt\nonumber\\
 &\sim \frac{A(B\gamma)^2 \Gamma(2\gamma -\alpha +1) }{|\Gamma(-\alpha)| (2+\alpha  - 2\gamma)} \frac{1}{s^{2\gamma-\alpha+1}}. 
\tag{S6}
\end{align}
Using Eq.~(S2) and 
\begin{align}
\hat{\psi}''(0,s) 
\sim \frac{AB^2 \Gamma(2\gamma-\alpha)}{|\Gamma(-\alpha)|}\frac{1}{s^{2\gamma-\alpha}}, \quad (\gamma >\alpha)
\tag{S7}
\end{align}  
we have
\begin{align}
\langle X^2_s \rangle &\sim \left[\frac{(B\gamma)^2 \Gamma(2\gamma -\alpha +1) }{|\Gamma(-\alpha)| (2+\alpha  - 2\gamma)} 
+ \frac{2B \Gamma(\gamma-\alpha)}{|\Gamma(-\alpha)|}\frac{B\gamma \Gamma(\gamma -\alpha +1) }{|\Gamma(-\alpha)| (1+\alpha  -\gamma)}
+ \frac{2B^2 \Gamma(\gamma-\alpha)^2}{|\Gamma(-\alpha)|^2}
+ \frac{B^2 \Gamma(2\gamma-\alpha)}{|\Gamma(-\alpha)|} \right]\frac{1}{s^{2\gamma+1}}\nonumber\\
&= \frac{B^2 }{|\Gamma(-\alpha)|}
\left[\frac{\gamma^2 (2\gamma-\alpha) \Gamma(2\gamma -\alpha) }{2+\alpha  - 2\gamma} 
+ \frac{2 \gamma (\gamma-\alpha) \Gamma(\gamma-\alpha)^2 }{|\Gamma(-\alpha)| (1+\alpha  -\gamma)}
+ \frac{2 \Gamma(\gamma-\alpha)^2}{|\Gamma(-\alpha)|}
+ \Gamma(2\gamma-\alpha) \right]\frac{1}{s^{2\gamma +1}}\nonumber\\
&= \frac{B^2 }{|\Gamma(-\alpha)|}
\left[\left\{1+\frac{\gamma^2 (2\gamma-\alpha)  }{2+\alpha  - 2\gamma}\right\} \Gamma(2\gamma -\alpha)
+ \left\{ 1 + \frac{\gamma (\gamma-\alpha) }{1+\alpha  -\gamma}
\right\} \frac{2 \Gamma(\gamma-\alpha)^2}{|\Gamma(-\alpha)|}
 \right]\frac{1}{s^{2\gamma +1}}.
 \tag{S8}
\end{align}
The inverse Laplace transform reads
\begin{align}
\langle X^2_t \rangle &\sim 
\frac{B^2 }{|\Gamma(-\alpha)|\Gamma(1+2\gamma)}
\left[\left\{1+\frac{\gamma^2 (2\gamma-\alpha)  }{2+\alpha  - 2\gamma}\right\} \Gamma(2\gamma -\alpha)
+ \left\{ 1 + \frac{\gamma (\gamma-\alpha) }{1+\alpha  -\gamma}
\right\} \frac{2 \Gamma(\gamma-\alpha)^2}{|\Gamma(-\alpha)|}
 \right] t^{2\gamma}.
\tag{S9}
\end{align}

\section{$n$-th ($n>1$) moment and its coefficient $M_n(\alpha,\gamma)$}

The $n$-th {($n>1$)} differentiation of $\hat{P}(k,s)$ is given by the recursion relation:
\begin{align}
\hat{P}^{(n)}(k,s) = \frac{1}{1-\hat{\psi}(k,s)} \left[ \sum_{i=1}^{n-1} c_{n,i}\hat{P}^{(i)}(k,s) \hat{\psi}^{(n-i)}(k,s) + 
\hat{P}(k,s) \hat{\psi}^{(n)}(k,s) + \int_0^\infty d\tau w(\tau) \hat{\Psi}^{(n)}(k,s;\tau)
\right],
\tag{S10}
\end{align}
where $c_{n,i}=c_{n-1,i}+c_{n-1,i-1}$ ($i=2, \ldots, n-2$) and $c_{n,n-1}=c_{n,1}=n$.
Here, 
\begin{align}
\int_0^\infty d\tau w(\tau) \hat{\Psi}^{(n)}(0,s;\tau) &= \int_0^\infty d\tau w(\tau) \int_0^\tau I(t,\tau)^n e^{-st}dt \nonumber\\
&= \int_0^\infty dt \int_t^\infty d\tau w(\tau) I(t,\tau)^n e^{-st} \nonumber\\
&\sim \frac{A(B\gamma)^n}{|\Gamma(-\alpha)|} \int_0^\infty dt t^n \int_t^\infty \tau^{-(n+1) -\alpha +n \gamma} d\tau  e^{-st} \nonumber\\
&\sim \frac{AB^n \gamma^n \Gamma (n\gamma -\alpha +1)}{|\Gamma(-\alpha)|(n+\alpha -n\gamma)}\frac{1}{s^{1-\alpha + n\gamma}}.
\tag{S11}
\end{align}
We assume 
\begin{align}
\hat{P}^{(i)}(0,s) \sim {(-1)^i}\frac{B^i M_i(\alpha,\gamma)}{|\Gamma(-\alpha)|} \frac{1}{s^{1+i\gamma}},
\tag{S12}
\end{align}
{for $i<n$.}
It follows that 
\begin{align}
\hat{P}^{(n)}(0,s) &=  \left[\sum_{i=1}^{n-1} c_{n,i} \frac{M_i(\alpha,\gamma)}{|\Gamma(-\alpha)|}  
\frac{\Gamma((n-i)\gamma - \alpha)}{|\Gamma(-\alpha)|} + 
\frac{\Gamma(n\gamma - \alpha)}{|\Gamma(-\alpha)|}
+  \frac{\gamma^n\Gamma(n\gamma - \alpha +1)}{|\Gamma(-\alpha)|(n+\alpha-n\gamma)} \right] \frac{(-B)^n}{s^{1 + n\gamma}} \nonumber\\
&= \left[\sum_{i=1}^{n-1} c_{n,i} M_i(\alpha,\gamma) \frac{\Gamma((n-i)\gamma - \alpha)}{|\Gamma(-\alpha)|} 
+  \left\{1+ \frac{\gamma^n(n\gamma-\alpha)}{n+\alpha-n\gamma}\right\} \Gamma(n\gamma - \alpha) \right] \frac{(-B)^n}{|\Gamma(-\alpha)| s^{1 + n\gamma}}. 
\tag{S13}
\end{align}
Therefore, 
\begin{align}
M_n (\alpha,\gamma) = \sum_{i=1}^{n-1} c_{n,i} M_i(\alpha,\gamma) \frac{\Gamma((n-i)\gamma - \alpha)}{|\Gamma(-\alpha)|} 
+ \left\{1+ \frac{\gamma^n(n\gamma-\alpha)}{n+\alpha-n\gamma}\right\}\Gamma(n\gamma - \alpha).
\tag{S14}
\end{align}

\bibliographystyle{apsrev} 

\end {document}